\documentstyle{mn}
\input epsf

\begin{document}

\newcommand{\Lya}{Ly$\alpha\ $}
\newcommand{\HI}{\hbox{H~$\scriptstyle\rm I\ $}}
\newcommand{\NHI}{$N_{\rm HI}\, $}
\newcommand{\rchi}{\chi_{\rm red}^2\, }
\newcommand{\hMpc}{$h^{-1}\, {\rm Mpc}\, $}
\newcommand{\Mpch}{$h\, {\rm Mpc}^{-1}\, $}
\newcommand{\kms}{${\rm km\, s}^{-1}\, $}

\newcommand{\etal}{~et al.\ }

\journal{Mon. Not. R. Astron. Soc. {\bf 000}, 1--8 (2000)}

\title{Spectral Analysis of the Ly$\alpha$ Forest Using Wavelets }

\author[A.Meiksin]{A. Meiksin \\
Institute for Astronomy, University of Edinburgh,\\
Blackford Hill, Edinburgh EH9 3HJ, UK}

\date{Accepted 1999 December 17. Received 1999 July 26}

\maketitle

\begin{abstract}
It is shown how wavelets may be used to analyse the absorption properties
of the \Lya forest. The Discrete Wavelet Transform of a QSO spectrum
is used to decompose the light fluctuations that comprise the forest into
orthogonal wavelets. It is demonstrated that most of the signal is carried
by the moderate to lower frequency wavelets in high resolution spectra, and
that a statistically acceptable description of even high signal--to--noise
spectra is provided by only a fraction (10--30\%) of the wavelets. The
distributions of the wavelet coefficients provide a statistical basis for
discriminating between different models of the \Lya forest.
The method is illustrated using the measured spectrum of Q1937--1009. The
procedure described is readily automated and may be used to process both
measured spectra and the large number of spectra generated by numerical
simulations, permitting a fair comparison between the two.

\end{abstract}


\section{Introduction}

Measurements of QSO spectra show that the Intergalactic Medium (IGM) is
composed of highly inhomogeneous structures. Ever since their identification
by Lynds (1971) and the pioneering survey of Sargent \etal (1980), these
inhomogeneities have been described as discrete absorption systems, the \Lya
forest. With the view that the systems arise from individual intervening gas
clouds, the \Lya forest has been characterized using traditional absorption
line statistics, most notably the line equivalent widths and, as the spectra
improved in resolution and signal--to--noise ratio, the Doppler widths and \HI
column densities through Voigt profile line fitting to the features.

In the past few years, numerical simulations have successfully modelled many
of the measured properties of the forest, showing that the absorption systems
may arise as a consequence of cosmological structure formation
(Cen \etal 1994; Zhang, Anninos \& Norman 1995; Hernquist \etal 1996;
Bond \& Wadsley 1997; Zhang \etal 1997; Theuns, Leonard \& Efstathiou 1998).
The simulations have shown, contrary to the picture in which the systems are
isolated intergalactic gas clouds, that most of the systems originate in an
interconnected web of sheets and filaments of gas and dark matter
(Cen \etal 1994; Bond \& Wadsley 1997; Zhang \etal 1998). Alternative
statistical methods were subsequently
introduced for describing the forest using the more direct measurements of the
induced light fluctuations. These include the 1-point distribution of the
fluctuations (Miralda-Escud\'e \etal 1996; Zhang \etal 1997), and a quantity
related to the 2-point distribution based on a weighted difference of the light
fluctuations in neighbouring wavelength pixels (Miralda-Escud\'e \etal).
A direct estimate of the 2-point transmission correlation function was made by
Zuo \& Bond (1994).

While the newer methods for analysing the \Lya forest avoid the identification
of absorption lines and the fitting of Voigt profiles, they are not
necessarily fundamentally different in their description of
the spectra. For instance, Zhang \etal (1998) find that the distribution of
optical depth per pixel in their simulation may be recovered by modelling the
spectra entirely by discrete absorption lines with Voigt profiles. Rather the
more direct methods circumvent a difficulty that has long plagued attempts to
characterize the absorbers in terms of Voigt profiles: the sensitivity of the
resulting line statistics to noise and to the fitting procedure. Absorption
line fitting of necessity requires arbitrary decisions to be made regarding
the setting of the continuum level, the deblending of features, and a decision
on the acceptability of a fit. Different observational
groups report different distributions for the line parameters. Most
discrepant has been the inferred distribution of line widths. Even with the
highest quality data gathered to date using the Keck HIRES, agreement is still
lacking, with Hu \etal (1995) finding a narrower Doppler parameter distribution
with a significantly higher mean than found by Kirkman \& Tytler (1997).
The differences are important, as cosmological simulations predict comparable
differences for a range of plausible cosmological models (Machacek \etal 2000;
Meiksin \etal 2000).

The purpose in this paper is to develop a method that provides an alternative
objective description of the statistics of the \Lya forest. Ultimately the
goal is to employ the same method for analysing both observational data and
data derived from numerical simulations in order to compare the two on a fair
basis. Because of the large number of synthetic spectra generated from a
simulation necessary to provide a correct average description of the forest,
two principal requirements of the procedure are that it be fast and easily
automated. Although automated or semi--automated Voigt profile fitting
procedures exist (AutoVP, Dav\'e \etal 1997; VPFIT, developed by Carswell and
collaborators), these procedures still
require arbitrary decisions to be made to obtain acceptable fits. The
complexity of the codes makes it difficult to assess the statistical
significance of differences between the measured distributions of the
absorption line parameters and those predicted. The codes also are
computationally expensive, making very costly their application to the large
number of simulated spectra required to obtain a statistically valid average
of the line parameters. For these reasons, a faster less complex method would
be desirable. The Voigt profile fitting codes yield important parameters, like
the linewidths, which contain physical information (eg, gas temperature and
turbulent velocities), that the direct-analysis methods do not. It
would thus be desirable for an alternative method to retain some of this
information. The method presented here utilizes wavelets to characterize the
absorption statistics of the \Lya forest. It is not intended to be a
replacement for Voigt profile fitting, but a fast alternative that allows
a ready comparison between the predictions of numerical models and measured
spectra and a clear statistical analysis of the results.

The outline of the paper is as follows: in \S\ref{sec:wavelets} it is shown
how the statistics of the \Lya forest may be characterized using wavelets.
In \S\ref{sec:results} the method is applied to the measured spectrum of a
high redshift QSO. The results are summarized in \S\ref{sec:summary}.

\section{Analysing the \Lya Forest with Wavelets} \label{sec:wavelets}

\subsection{Terminology} \label{sec:terms}

Although wavelets have been used in signal processing, image analysis, and the
study of fluid dynamics for a decade, they are only beginning to enter the
vernacular of astronomers. Accessible introductions are provided in
Press \etal (1992), and in Slezak, Bijaoui \& Mars (1990) and
Pando \& Fang (1996), who apply wavelets to study the clustering of galaxies
and \Lya absorbers, respectively. More complete accounts of wavelet
methodology are Chui (1992), Daubechies (1992), and Meyer (1993). The
description here is confined to those elements necessary to introduce the
notation and terminology that will be used below.

Wavelets are defined variously in the literature. The definition of most use
here, somewhat restrictive but appropriate to a multiresolution analysis
using the Discrete Wavelet Transform (DWT), is (Meyer):

\begin{quote}
A {\it wavelet} is a square--integrable function $\psi(x)$ defined in real
space such that $\psi_{jk}\equiv2^{j/2}\psi(2^jx-k)$, where $j$ and $k$ are
integers, is an orthonormal basis for the set of square--integrable functions.
\end{quote}
The wavelet $\psi(x)$ satisfies $\int_{-\infty}^{\infty} dx\, \psi(x)=0$,
and is generally chosen to be concentrated near $x=k2^{-j}$.
Its defining properties permit it to perform two operations
governed by the values of $j$ and $k$. Smaller values of $j$ correspond to
coarser variations in $f(x)$, while differing values of $k$ correspond to
shifting the centre of the transform.

The {\it wavelet coefficients} of a function $f(x)$ are defined by
\begin{equation}
w_{jk}\equiv\int dx\, f(x)\psi_{jk}(x).
\end{equation}
The set of coefficients $\{w_{jk}\}$
comprises the wavelet transform of the function $f(x)$. The function may then
be recovered through the inverse transform
\label{eq:wavelet_trans}
\begin{equation}
f(x) = \sum_{j,k} w_{jk}\psi_{jk}(x),
\label{eq:wavelet_inv}
\end{equation}
since the set of functions $\psi_{jk}$ forms a complete orthonormal basis.
The wavelet coefficients at a level $j$ express the changes between the
smoothed representations of $f(x)$ at the resolution scales $j+1$ and $j$.

Several functions may serve as wavelets. A set that has proven particularly
useful was developed by Daubechies (Daubechies 1992). These functions are
constructed to have vanishing moments up to some value $p$, and the functions
themselves vanish outside the range $0<x<2p+1$. The wavelet coefficients
decrease rapidly with $p$ for smooth functions. Accordingly, the higher order
Daubechies wavelets are the most suitable for analyzing smooth data. The DWT
is computed using the pyramidal algorithm as implemented in Numerical Recipes
(Press \etal). The Daubechies wavelet of order 20 is chosen throughout.

\subsection{Monte Carlo simulations} \label{sec:Monte}

The properties of the wavelet transform of the \Lya forest are examined by
performing Monte Carlo realizations of spectra. The spectra are constructed
from discrete lines with Voigt profiles using the \HI column density and
Doppler parameter distributions found by Kirkman \& Tytler. Specifically, the
\HI column densities \NHI are drawn from a power law distribution of slope 1.5
between $12.5<\log_{10}N_{\rm HI}<16$ and the Doppler parameters $b$ from a
gaussian with mean 23~\kms and standard deviation 14~\kms. A cut--off in $b$ is
imposed according to $b>14 + 4(\log_{10} N_{\rm HI} - 12.5)$~\kms. The
resulting average Doppler parameter is 31~\kms. The number density of lines
per unit redshift matches that of Kirkman \& Tytler at $z=3$. The resolution is
set at $\lambda/d\lambda=5\times10^4$, and gaussian noise is added according to
a specified continuum signal--to--noise ratio per pixel. This is the fiducial
model used in all the simulations unless stated otherwise. Segments 128 pixels
wide were found adequate for extracting the statistical properties of the
wavelet coefficients.

\begin{figure}
\centering
\epsfxsize=3.3in \epsfbox{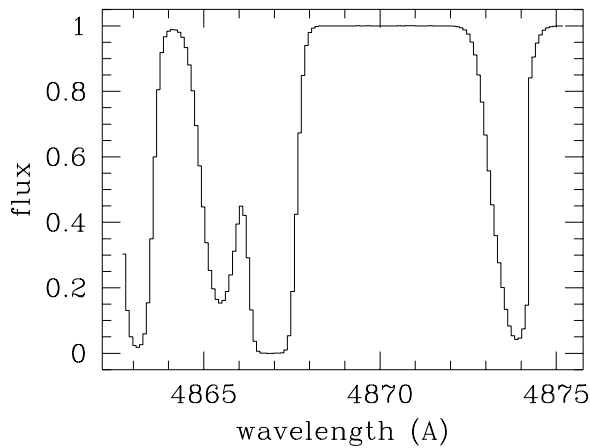}
\caption{(a)\ A representative synthetic spectrum showing the \Lya forest at
$z=3.0$ at a resolution of $\lambda/d\lambda=5\times10^4$. (b)\ The absolute
magnitudes of the wavelet coefficients are shown in the grayscale map. The map
is linear and ranges between 0 (white) and 0.5 (black).}
\label{fig:spec}
\end{figure}

A representative spectrum and its discrete wavelet transform are shown in
Figure~\ref{fig:spec}. A block at resolution $j$ is $128/2^j$ pixels wide and
$2^j$ pixels long for $j=1$ to 6. The resolution becomes finer as $j$ increases
from 1 to 6 (downwards). The uppermost level ($j=0$) corresponds to smoothed
averages of the spectrum. The wavelet coefficients tend to increase in
magnitude with decreasing resolution (decreasing $j$). The low values indicate
that only small changes occur in the spectrum when smoothed at one resolution
level to the next higher. The small values are desirable, as they signify
the dominant absorption features in the spectra are adequately resolved. 

Because the wavelet functions form a complete set of basis functions,
the full set of wavelet coefficients completely describes the
spectrum: the spectrum may be reconstructed identically from the
inverse transform. For noisy spectra, however, it will generally be
unnecessary to retain the full set of coefficients.  Indeed, this is
the motivation for multi--resolution data compression. By
employing a judicious set of basis functions, a signal may be
compressed into only a small fraction of its original size. The method
of chosing the optimal basis set such that the compressed signal matches the
original as closely as possible in a least squares sense with the least
number of retained basis elements is known as Proper
Orthogonal Decomposition or the Karhunen--Lo\'eve procedure (see Berkooz,
Holmes \& Lumley 1993 for a review). The basis set, however, will in general
differ from signal to signal if its components are highly variable, as in the
case of the \Lya forest. Although not optimal in the least squares sense,
the wavelet basis nonetheless achieves a large amount of data compression and
has the advantage of generality. Next is described how wavelets may be applied
to assessing the amount of useful information in a spectrum.

Two measures of the information content of a noisy spectrum are considered,
one based on $\chi^2$ and the second on entropy. If $s(x_i)$ is the original
spectrum defined at $N$ points $x_i$ (eg, wavelength or velocity), and
$s_n(x_i)$ is the spectrum reconstructed from the $n$ largest (in magnitude)
wavelet coefficients, then
\begin{equation}
\chi^2=\sum_{i=1}^{N}\left[\frac{s(x_i)-s_n(x_i)}{\sigma_i}\right]^2
\label{eq:chi2}
\end{equation}
where $\sigma_i$ is the measurement error associated with pixel $i$. For
gaussian distributed measurements, the expectation value of $\chi^2$ is
the number of degrees--of--freedom. If $n$ wavelet coefficients are retained,
the number of degrees--of--freedom is $N-n$. (Hence, for example, $\chi^2=0$
is expected for $n=N$.) The reduced $\rchi=\chi^2/(N-n)$ then
defines the optimal value of $n$ for truncating the wavelet coefficients.

The information content may also be expressed in terms of the wavelet
coefficients directly as an ``entropy''\footnote{Meyer (1993)
defines the entropy to be the exponential of $S$.}
\begin{equation}
S=-\sum_{jk}\, \alpha^2_{jk}\log \alpha^2_{jk},
\label{eq:entropy}
\end{equation}
where the $\alpha_{jk}$ are the normalized coefficients
\begin{equation}
\alpha_{jk}=\frac{w_{jk}}{\left(\sum_{jk} w^2_{jk}\right)^{1/2}}.
\label{eq:alphajk}
\end{equation}
This quantity behaves like a physical entropy in the sense that it is
maximum when the signal is completely random so that the full set of
coefficients $\{w_{jk}\}$ is required to describe it, while it vanishes
when the signal may be entirely described by a single coefficient.

\begin{figure}
\begin{center}
\leavevmode \epsfxsize=3.3in \epsfbox{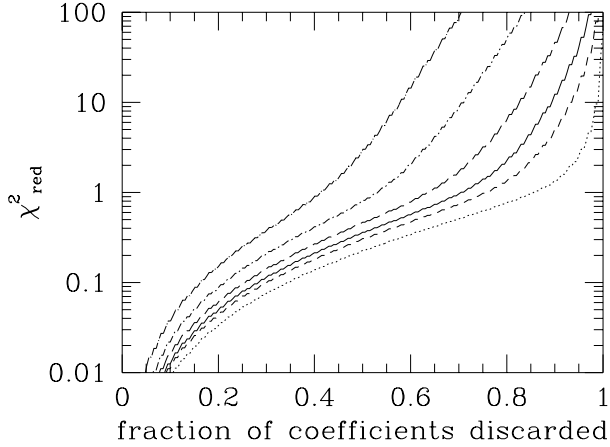}
\end{center}
\caption{The dependence of the reduced $\rchi$ on the fraction
of discarded wavelet coefficients. The curves increasing from the bottom
are for signal--to--noise ratios of 10, 30, 50, 100, 300, and 1000.
As the noise level increases, an increasing
fraction of the coefficients may be discarded with the remainder still
providing a statistically acceptable fit to the spectra.}
\label{fig:chi2}
\end{figure}

The reduced $\chi^2$ for an ensemble of Monte Carlo realizations is shown in
Fig.~\ref{fig:chi2} as a function of the fraction $(N-n)/N$ of the wavelet
coefficients discarded. As the signal--to--noise ratio per pixel increases,
the value of $\rchi$ for a given $n$ increases. In all cases,
however, there is some $n<N$ for which $\rchi=1$. This suggests
that an acceptable fit to a noisy spectrum may be provided by only a fraction
$n/N$ of the full set of coefficients, with the fraction required increasing as
the noise level decreases.

\begin{figure}
\begin{center}
\leavevmode \epsfxsize=3.3in \epsfbox{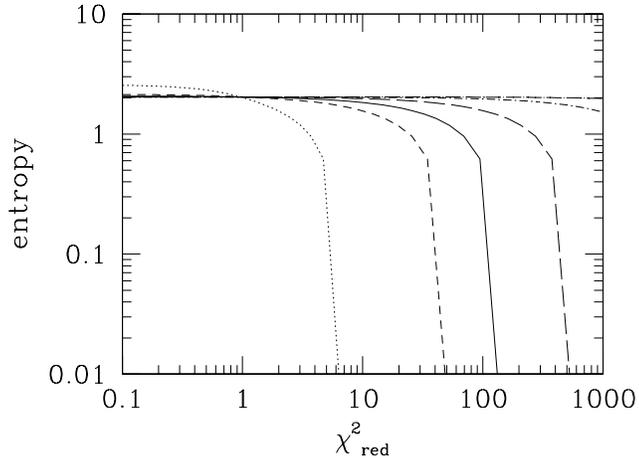}
\end{center}
\caption{The entropy of the spectra defined in
terms of the wavelet coefficients. Little information is lost from the
spectra for $\rchi\le1$, as measured by the entropy. The entropy curves from
left to right are for signal--to--noise ratios of 10, 30, 50, 100, 300, and
1000.}
\label{fig:entropy}
\end{figure}

The entropy $S$ is shown in Figure~\ref{fig:entropy}.
The entropy stays nearly constant out to $\chi_{\rm red}^2=1$,
indicating that little information has been lost by discarding the small
coefficients. As $\rchi$ increases, eventually the entropy decreases
as information is lost. Due to the greater information content of the less
noisy spectra, as the noise level is decreased, the entropy remains constant
to increasingly higher values of $\rchi$ before declining.

\subsection{Statistics of the wavelet coefficients}

It was shown above how wavelets may be used to characterize the noise
properties of a spectrum. The wavelet coefficients, however, may also be
used to characterize the statistics of the \Lya forest itself.

\begin{figure}
\begin{center}
\leavevmode \epsfxsize=3.3in \epsfbox{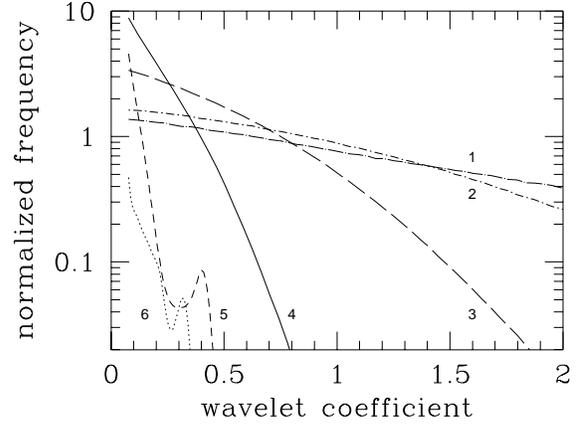}
\end{center}
\caption{The normalized distribution of the wavelet coefficients at the levels
$j=1\dots6$. The coefficients increase in magnitude at the coarser
(lower $j$) resolutions, indicating that they carry most of the information
in the spectra.}
\label{fig:wcdist}
\end{figure}

The distributions of the coefficients (in absolute value) for the several
resolution levels are shown in Fig.~\ref{fig:wcdist} for a set of simulated
spectra with $S/N=50$, typical of the Keck HIRES spectra. The number of
coefficients at a level $j$ is $2^j$, with $j=1$ corresponding to the
coarsest resolution, and $j=6$ to the finest for the $2^7=128$ pixels used in
a spectrum. The finest resolution ($j=6$) curve is the steepest. As the
resolution becomes increasingly coarse, the amplitude of the coefficients
increases, as was found in Fig.~\ref{fig:spec}. This indicates that most of
the information in the spectrum is carried by the coarser levels (as well as
by the two course scale averages, not shown). The finest level has
resolved the spectral structures, with little difference between the smoothed
representations of the spectrum at resolution levels $j=5$ and 6.
Applying a cut--off in the coefficients corresponding to
$\rchi=1$ yields for the average number retained of the initially
$2^j$ coefficients for $j=1\dots6$ the respective values 1.9, 3.8, 7.4, 11.6,
6.9, and 3.4. While almost all of the coefficients for $j\le3$ are needed,
a decreasing fraction is required to describe the spectra at higher resolution.

\begin{figure}
\begin{center}
\leavevmode \epsfxsize=3.3in \epsfbox{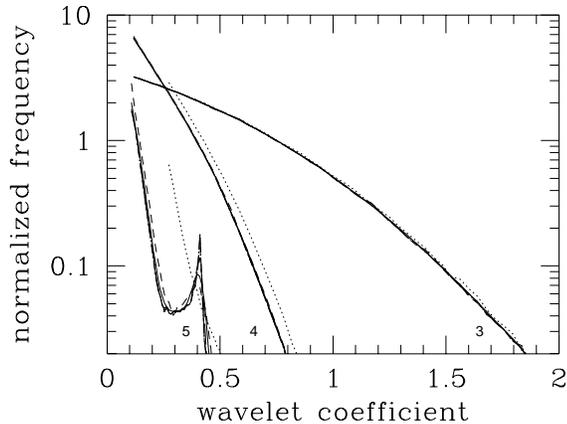}
\end{center}
\caption{The normalized distribution of the wavelet coefficients for the levels
$j=3, 4,$ and 5 for signal--to--noise ratios of 10, 30, 50, 100, 300, and
1000. Except for $S/N=10$ (dotted lines), the distributions overlap.}
\label{fig:dist_snr}
\end{figure}

The distributions are insensitive to the signal--to--noise ratio, as shown
in Fig.~\ref{fig:dist_snr}. Except for the lowest ratio of 10, the curves
coincide, showing that they may be measured accurately even for a varying
signal--to--noise ratio in a spectrum, provided it is not too low.

\begin{figure}
\begin{center}
\leavevmode \epsfxsize=3.3in \epsfbox{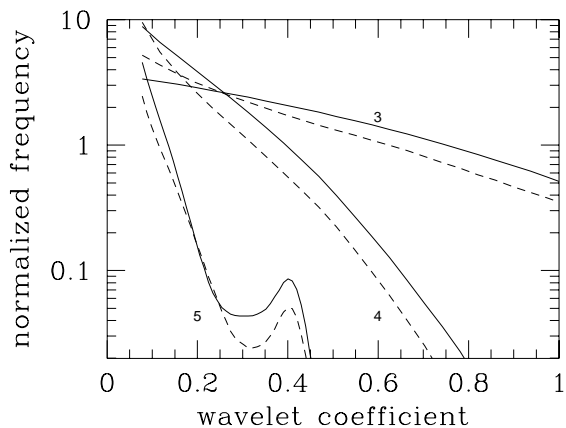}
\end{center}
\caption{The normalized frequencies of the wavelet coefficients for the levels
$j=3, 4,$ and 5 for two different statistical descriptions of the \Lya forest.
The solid lines are based on the Voigt parameter distributions inferred by
Kirkman \& Tytler, and the dashed on the distributions inferred by Hu \etal
A signal--to--noise ratio of 50 is used in both sets of simulations. The
distributions of wavelet coefficients distinguish between the two models.}
\label{fig:dist_KTH}
\end{figure}

To demonstrate that the wavelet coefficient distributions may be used to
discriminate between different predictions for the statistical properties
of the \Lya forest, a second set of Monte Carlo realizations with alternative
column density and Doppler parameter distributions is generated. The
parameters adopted are those reported by Hu \etal They found that the forest
statistics are consistent with an \HI column density distribution with a slope
of $1.5$ for clouds with $12.3<\log N_{\rm HI}<14.5$ and a Gaussian Doppler
parameter distribution with mean 28~\kms, standard deviation 10~\kms, and
a sharp cut--off below 20~\kms. The resulting average Doppler parameter is
37~\kms. The simulation is normalized to the same line density per unit
redshift at $z=3$ as found by Hu \etal, but the column density distribution
is extended to $\log N_{\rm HI}=16$ to be consistent with the previous set of
simulations. The resulting wavelet coefficient distributions are compared with
those from the previous simulation for $j=3,4,$ and 5 in
Fig.~\ref{fig:dist_KTH}. The wavelet coefficients are able to distinguish
between the two models.

The Kolmogorov--Smirnov test may be used to assess the probability that the
wavelet coefficients of a measured spectrum match a given distribution for each
resolution level $j$. The most stringent test, however, is given by combining
the probabilities for all the distributions. Because any given absorption
feature may be expected to affect the coefficients at more than a single
resolution level $j$, it is possible that the coefficients corresponding
to a given set of nested blocks for different $j$ (see Fig.~\ref{fig:spec})
may be correlated. In this case, the probabilities of matching the various
distributions may not be combined as if they were independent. To determine
the degree to which the distributions may be treated as independent, the
correlations are measured for coefficients between the various levels $j$
corresponding to the same hierarchy of blocks, and then averaged over all the
hierarchies, for a set of Monte Carlo
realizations using the fiducial forest model. The results  are shown in
Table~\ref{tab:corr}. (The level $j=0$ refers to the correlations with the
pair of coefficients corresponding to the course scale averages.)
A signal--to--noise ratio of 50 is assumed, and a cut-off in the coefficients
is applied to ensure $\rchi=1$. The error on the correlations is
$\sim0.1$\%. Although the correlations are small, they are not absent. They
are sufficiently small, however, that treating the probabilities for the
different distributions as independent should be an adequate approximation
for model testing.

\begin{table*}
\caption{Wavelet coefficient correlation matrix for resolution levels
$j=6,\dots,0$.}
\begin{tabular}{lrrrrrrr}
$j$   & 6 & 5 &  4 & 3 & 2 & 1 & 0 \\
\\
6 &  1.000 & 0.003 & 0.009 & 0.013 & 0.009 & 0.003 & 0.001 \\
5 &  0.003 & 1.000 & -0.020 & -0.013 & -0.005 & -0.001 & -0.001 \\
4 &  0.009 & -0.020 & 1.000 & 0.026 & 0.018 & 0.012 & 0.000 \\
3 &  0.013 & -0.013 & 0.026 & 1.000 & 0.046 & 0.025 & 0.002 \\
2 &  0.009 & -0.005 & 0.018 & 0.046 & 1.000 & 0.035 & 0.005 \\
1 &  0.003 & -0.001 & 0.012 & 0.025 & 0.035 & 1.000 & -0.002 \\
0 &  0.001 & -0.001 & 0.000 & 0.002 & 0.005 & -0.002 & 1.000 \\
\end{tabular}
\label{tab:corr}
\end{table*}

\subsection{Data compression} \label{sec:compress}

One of the key features of wavelets is their ability to compress data.
Figs.~\ref{fig:chi2} and \ref{fig:entropy} show that it is possible to fit a
spectrum using only a subset of the wavelets used in its DWT at a statistically
acceptable level ($\rchi=1$), without significantly degrading the
information content of the spectrum as measured by the wavelet entropy.
This suggests that filtering the spectrum in this way may provide a usable
spectrum that is relatively noiseless and suitable for absorption line fitting.

\begin{figure}
\begin{center}
\leavevmode \epsfxsize=3.3in \epsfbox{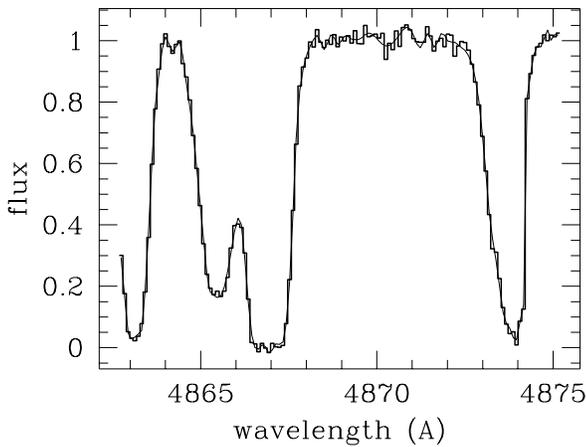}
\end{center}
\caption{Synthetic spectrum with $S/N=50$ (heavy solid histogram). The wavelet
filtered spectrum with $\rchi=1$ is shown by the lighter
line (shown as a smooth curve for clarity).}
\label{fig:specwfit}
\end{figure}

This is illustrated by performing Voigt profile fitting to Monte Carlo
realisations of the fiducial line model, with an assumed signal--to--noise
ratio of 50. A wavelet filtered representation of each realised spectrum is
generated with coefficients truncated to give a reduced $\rchi=1$ for the
difference between the original and wavelet filtered spectra. This
corresponds on average to retaining only 30\% of the full set of coefficients.
A representative spectrum is shown in Fig.~\ref{fig:specwfit}.

\begin{figure}
\begin{center}
\leavevmode \epsfxsize=3.3in \epsfbox{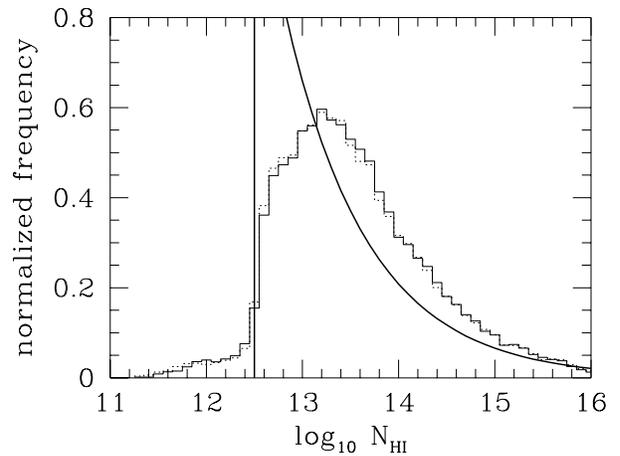}
\end{center}
\caption{The recovered \HI column density distribution from a set of Monte
Carlo realizations. The dotted curve corresponds to the Voigt fits obtained
using the original unfiltered spectra. The solid curve shows the recovered
distribution obtained from the wavelet filtered spectrum for which only 30\%
of the wavelet coefficients are retained, corresponding to a reduced $\rchi=1$
for the difference between the original and wavelet filtered spectra. The heavy
solid line shows the input model distribution.}
\label{fig:NHIdist}
\end{figure}

\begin{figure}
\begin{center}
\leavevmode \epsfxsize=3.3in \epsfbox{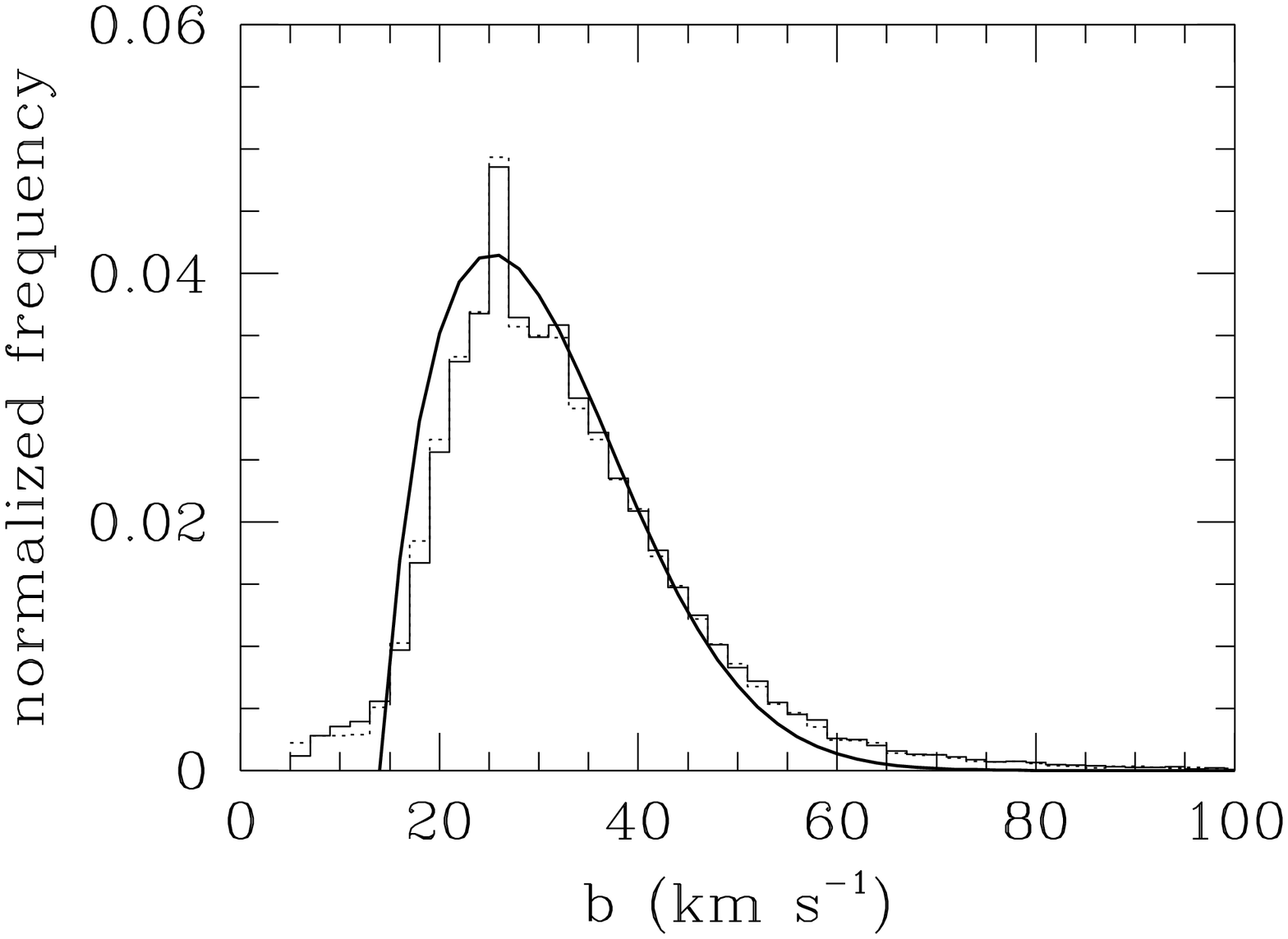}
\end{center}
\caption{The recovered Doppler parameter distribution, as in
Fig.~\ref{fig:NHIdist}. The heavy solid line shows the input model
distribution.}
\label{fig:bdist}
\end{figure}

Absorption lines are then identified in the filtered spectrum and fit
using AutoVP. The results of $10^4$ realisations are shown in
Figs.~\ref{fig:NHIdist} and \ref{fig:bdist}. Also shown are the
distributions obtained from AutoVP using the original spectra with no
wavelet filtering applied. The distributions are nearly identical. A
negligible loss is incurred in the recovery of the line parameters
despite the exclusion of 70\% of the information in the original spectra.

\section{Application to Q1937--1009} \label{sec:results}

In this section, the Discrete Wavelet Transform is used to analyse the \Lya
forest as measured in the $z=3.806$ QSO Q1937--1009. The spectrum was taken
with the Keck HIRES at a resolution of $\sim8.5$~\kms (Burles \& Tytler 1997).
The signal--to--noise ratio per pixel was $\sim50$. The spectrum covers the
range between \Lya and Ly$\beta$ in the QSO restframe. (The region analysed
is restricted to the redshift interval $3.055<z<3.726$ to avoid any possible
influence by the QSO.)

\begin{figure}
\begin{center}
\leavevmode \epsfxsize=3.3in \epsfbox{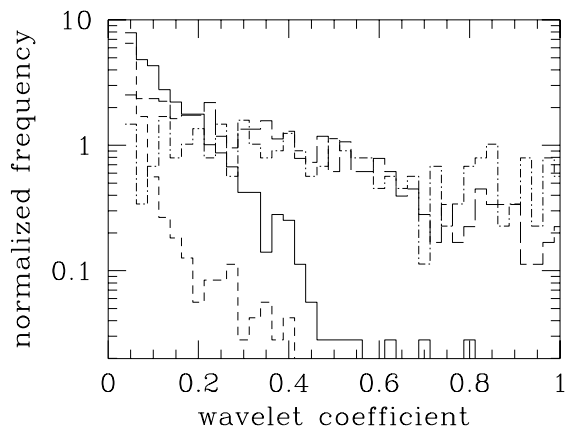}
\end{center}
\caption{The normalized distributions of the wavelet coefficients for the
spectrum of Q1937--1009. The distributions are shown for the levels
$j=2$ (dot--dashed), $j=3$ (long--dashed), $j=4$ (solid), and $j=5$
(short--dashed). The faster decline at higher frequencies demonstrates that
the absorption features dominating the spectrum have been adequately
resolved.}
\label{fig:wcKeck}
\end{figure}

The distribution of wavelet coefficients is shown in Fig.~\ref{fig:wcKeck} for
$j=2$, 3, 4, and 5. As in Fig.~\ref{fig:wcdist}, the high frequency
coefficients are generally smaller than at lower frequencies, indicating that
the fluctuations that dominate the spectra have been resolved.

\begin{figure}
\begin{center}
\leavevmode \epsfxsize=3.3in \epsfbox{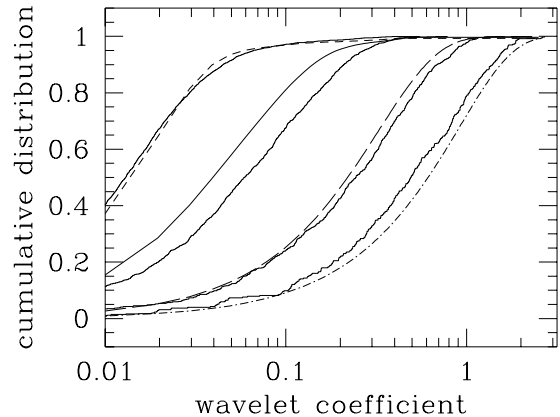}
\end{center}
\caption{The cumulative distributions of the wavelet coefficients for the
spectrum of Q1937--1009, along with the predicted distributions according
to the line model of Kirkman \& Tytler (1997). The frequency levels shown
are $j=2$ (dot--dashed), $j=3$ (long--dashed), $j=4$ (solid), and $j=5$
(short--dashed).}
\label{fig:wcumdKeck}
\end{figure}

The cumulative distributions of the coefficients are compared with the
predicted distributions for the fiducial model in Fig.~\ref{fig:wcumdKeck}.
The predicted distributions were generated by simulating spectra with
the same pixelization, resolution, signal--to--noise ratio and wavelength
coverage as for the measured spectrum of Q1937--1009. An increase in line
density per unit redshift proportional to $(1+z)^{2.6}$ (Kirkman \& Tytler)
was included to match to the redshift range of Q1937--1009. While the
distributions generally agree well, a large variation is found for $j=4$,
corresponding to fluctuations on the scale of $17-34$~\kms, suggesting some
differences from the line model of Kirkman \& Tytler. Effects neglected in
the simulations that could produce a difference are the presence of metal
systems and redshift correlations between the \Lya absorption systems. The
changes that would be produced, however, are most likely small:\ the number of
metal systems is small, and the correlations appear weak or absent (Meiksin \&
Bouchet 1995; Kim \etal 1997). Still, the sensitivity of the wavelet
coefficient distributions to these effects may be worth more careful
consideration.

\section{Summary} \label{sec:summary}

Wavelets may be usefully employed to provide a statistical
characterizaton of the absorption properties of the \Lya forest. An
approach is presented that performs a multiresolution analysis of the
forest using the Discrete Wavelet Transform of the QSO spectrum. The
transform decomposes the local frequency dependence of the light
fluctuations into an orthogonal hierarchy of basis functions, the
wavelets. It is found that in spectra of better than 10~\kms
resolution, most of the information of the spectrum is carried by
the lower frequency wavelets. For a signal--to--noise ratio typical of
even the highest quality spectra ($S/N=10-100$), only 10--30\% of the
wavelets are required to provide a statistically acceptable
description of the spectrum, corresponding to a data compression factor
of 3--10. It is shown that a Voigt profile line analysis performed on
the wavelet filtered spectra yields nearly identical line parameter
distributions as obtained from the original unfiltered spectra.

The distributions of the wavelet coefficients offer an alternative statistical
description of the \Lya forest while retaining information on the line widths.
It is demonstrated that the correlations of
coefficients between different levels in the wavelet hierarchy are weak (a few
percent or smaller). Consequently, each of the distributions may be treated as
statistically independent to good approximation.

The method is applied to a Keck HIRES spectrum of Q1937--1009. The
wavelet coefficient distributions behave qualitatively
similarly to those found in Monte Carlo simulations based on the line
parameter distributions reported by Kirkman \& Tytler. The measured
distributions, however, show some differences on the scale $17-34$~\kms.

The results demonstrate that Multiresolution Analysis using the
Discrete Wavelet Transform provides an alternative objective, easily
automated procedure for analysing the \Lya forest suitable for basing
a comparison between the measured properties of the \Lya forest and
the predictions of numerical models.

\bigskip
The author thanks S. Burles and D. Tytler for kindly providing the spectrum of
Q1937--1009, and R. Dav\'e for permission to use AutoVP.

\end{document}